\documentclass[letterpaper,12pt]{article}
\pdfoutput=1

\usepackage{jheppub}
\usepackage{hyperref}
\usepackage{amsmath}
\usepackage{amsfonts}
\usepackage{graphicx}
\usepackage{caption}
\usepackage{placeins}

\usepackage{comment}
\usepackage{array}
\usepackage{mathtools}
\def\be{\begin{equation}}
\def\ee{\end{equation}}
\def\ba{\begin{eqnarray}}
\def\ea{\end{eqnarray}}

\title{
Causal Diamonds, Cluster Polytopes and Scattering Amplitudes
} 

\author[a]{N.~Arkani-Hamed,}\author[b]{S.~He,}\author[c]{G.~Salvatori,}\author[d]{H.~Thomas}

\affiliation[a]{School of Natural Sciences, Institute for Advanced Study, Princeton, NJ, 08540, USA \\
Center of Mathematical Sciences and Applications, Harvard University, Cambridge, MA 02138, USA}
\affiliation[b]{%
CAS Key Laboratory of Theoretical Physics, Institute of Theoretical Physics, Chinese Academy of Sciences, Beijing 100190, China
\\
School of Fundamental Physics and Mathematical Sciences, Hangzhou Institute for Advanced Study, UCAS; International Centre for Theoretical Physics Asia-Pacific, Beijing/Hangzhou, China 
\\
Peng Huanwu Center for Fundamental Theory, Hefei, Anhui 230026, China
}
\affiliation[c]{Department of Physics, Brown University, Providence RI 02912, USA}
\affiliation[d]{LaCIM, D\'epartement de Math\'ematiques, Universit\'e du Qu\'ebec \`a Montr\'eal, Montr\'eal, QC, Canada}

\emailAdd{arkani@ias.edu}
\emailAdd{songhe@itp.ac.cn}
\emailAdd{giulio\_salvatori@brown.edu}

\date{\today}

\abstract{
The ``amplituhedron" for tree-level scattering amplitudes in the bi-adjoint $\phi^3$ theory is given by the ABHY associahedron in kinematic space, which has been generalized to give a realization for all finite-type cluster algebra polytopes, labelled by Dynkin diagrams. In this letter we identify a simple physical origin for these polytopes, associated with an interesting $(1+1)$-dimensional causal structure in kinematic space, along with solutions to the wave equation in this kinematic ``spacetime" with a natural positivity property. The notion of time evolution in this kinematic spacetime can be abstracted away to a certain ``walk", associated with any acyclic quiver, remarkably yielding a finite cluster polytope for the case of Dynkin quivers. The ${\cal A}_{n{-}3},{\cal B}_{n{-}1}/{\cal C}_{n{-}1}$ and ${\cal D}_n$ polytopes are the amplituhedra for $n$-point tree amplitudes, one-loop tadpole diagrams, and full integrand of one-loop amplitudes. We also introduce a polytope $\bar{\cal D}_n$, which chops the ${\cal D}_n$ polytope in half along a symmetry plane, capturing one-loop amplitudes in a more efficient way. 
}

\begin{document}
  \maketitle

\section{Scattering Forms and Positive Geometry in Kinematic Space}

Scattering amplitudes are  observables measured at the boundary of flat spacetime. They are determined solely by the data of ``kinematic space", corresponding to various ways of labelling the on-shell data of the scattering process. It is thus natural to ask: is there a question, directly posed in this kinematical space, whose answer yields the scattering amplitude, without invoking local evolution through the bulk of spacetime? 

The past few years have seen significant inroads in this program, beginning with the formulation of scattering amplitudes in planar ${\cal N}=4$ SYM, as a certain differential form in the kinematic momentum-twistor space~\cite{Hodges:2009hk}, fixed by pulling back on a family of subspaces, to the canonical form~\cite{Arkani-Hamed:2017tmz} of a kinematic-space avatar of the amplituhedron~\cite{Arkani-Hamed:2013jha, ArkaniHamed:2012nw}, which is in turn fully determined by natural notions of positivity and topology~\cite{Arkani-Hamed:2017vfh}. More recently, the same structure has been seen in the much simpler setting of tree amplitudes for the bi-adjoint $\phi^3$ theory~\cite{Arkani-Hamed:2017mur}. Again the amplitudes are naturally upgraded to a differential form on the kinematic space of Mandelstam invariants, determined by pulling back to the canonical form of a specific realization of the associahedron polytope~\cite{Stasheff_1, Stasheff_2} on a family of subspaces. This realization also exposes a hidden ``projective invariance" symmetry of the scattering form, analogous to the dual conformal invariance of ${\cal N}=4$ SYM~\cite{Drummond:2006rz,Drummond:2008vq}. This symmetry is invisible term-by-term in the diagrammatic expansion, but can be manifested by new representations of amplitudes, even for this seemingly simplest possible scalar theory with no supersymmetry or signs of integrable structure. 

The ABHY construction was soon generalized to all polytopes associated with finite-type cluster algebras~\cite{bazier2018abhy}, which are classified by Dynkin diagrams~\cite{fomin2003systems, chapoton2002polytopal}, with the usual associahedron corresponding to the simplest case of type ${\cal A}_n$ Dynkin diagrams. 

These developments have exposed structural aspects of particle scattering, which are hidden in the usual picture of local, unitary evolution through spacetime given by Feynman diagrams. Consider the most important feature of tree-level amplitudes: they have poles when the sum of a subset of external momenta goes on-shell, and the residue of the amplitude on this pole factorizes into a product of lower-point amplitudes. Indeed the purpose in life of the conventional picture of particle worldlines in space-time, as well as the string worldsheet, is to make factorization manifest. But there are {\it still further} simple, qualitative properties of the amplitude that are not captured by either the ``particle" or ``string" picture. This is seen in an even more basic and coarse characterization of the pole structure. We know that the tree amplitudes for $n$-particle scattering involve up to $(n-3)$ poles at a time, so it is interesting to ask about the pattern of poles that are ``compatible", {\it i.e.} that can appear together. Remarkably, this purely combinatorial question has a geometric answer: the sets of poles that appear together -- which can also be associated with the corresponding Feynman diagrams -- can be realized as the vertices of an $(n-3)$ dimensional polytope, the associahedron.
This is a highly non-trivial and surprising fact, which {\it also} explains factorization, since the facets of the associahedron can be seen to factorize into products of lower dimensional associahedra.

We can thus ask a sharpened version of our motivating question--is there a structure, living directly in the kinematic space -- that makes {\it both} the polytopal nature of the pole structure of the amplitudes, as well as factorization, completely obvious? Equivalently in our setting -- is there a question in kinematic space that makes the ABHY associahedra natural and inevitable?

This is the challenge we take up in this letter. We will begin by seeing that the kinematical variables associated with tree amplitudes are naturally associated with a $(1+1)$ dimensional ``kinematic spacetime" geometry. We then ask a simple question in this spacetime, looking for positive solutions of the wave equation with positive source. Remarkably, we find that the space of solutions are generalized ABHY associahedra. The positive wave equation picture makes the existence of these polytopes, as well as its factorization properties, completely obvious, following directly from simple properties of causal diamonds in the spacetime. 

It is also natural to solve the wave equation via time evolution on slices through the $(1+1)$ dimensional kinematic spacetime. We will learn to describe this time evolution in a somewhat more abstract way, in terms of a simple set of ``mutations" on a quiver diagram naturally associated with time slices through this spacetime. This will allow us define a generalized notion of ``time evolution", associated with completely general quivers. We can then ask when following this rule produces a finite polytope. Remarkably, we find that this happens when the quivers are Dynkin diagrams,  and the corresponding polytopes are the generalized ABHY associahedra of \cite{bazier2018abhy} for all the finite-type cluster algebras. The notion of ``time evolution" in this case reproduces the well-known ``Auslander-Reiten quiver" walk through all cluster variables of finite-type cluster algebras used in the construction of \cite{bazier2018abhy}. The polytopes constructed in this way  factorize on their boundaries, to the product of the polytopes associated with the Dynkin diagrams obtained by removing a node from the original one.

\begin{figure}[h!]
    \centering
    \includegraphics[scale=1]{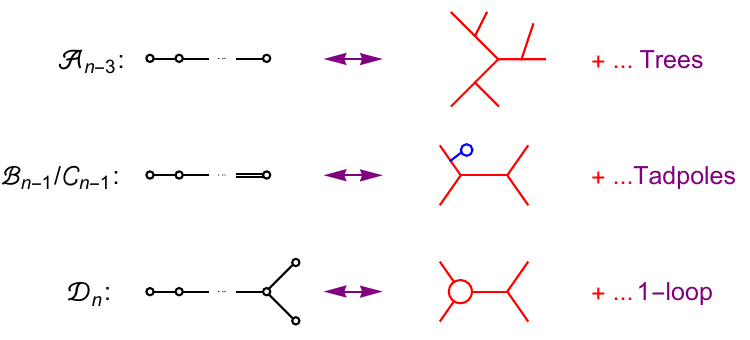}
    \caption{\label{fig:algAndGraph} The correspondence between cluster algebras, Dynkin quivers and scattering amplitudes}
\end{figure}
\FloatBarrier

{ As shown in Figure \ref{fig:algAndGraph}, the polytopes associated with the classical ${\cal A},{\cal B}/{\cal C}, {\cal D}$ Dynkin quivers are especially interesting in connection with physics}. We already know that ${\cal A}_{n{-}3}$ produces $n$-particle tree amplitudes in bi-adjoint $\phi^3$ theory. The factorization behavior of ${\cal D}_n$  is exactly what is needed for this polytope to produce $n$-particle 1-loop amplitudes. We also identify a smaller polytope $\bar{\cal D}_n$ by chopping ${\cal D}_n$ in half along a $\mathbb{Z}_2$ symmetry plane, which gives 1-loop amplitudes in a more efficient way. Similarly the polytopes for ${\cal B}_{n{-}1}/{\cal C}_{n{-}1}$, which turn out to be the same and are known as ``cyclochedra", compute the subset of 1-loop amplitudes involving the emission of tadpoles. This polytope is also directly realized as one of the facets of the $\overline{\cal D}_n$ polytope. 

Our aim in this letter is to highlight this striking connection between causal diamonds in kinematic space and cluster polytopes, and to give an explicit description of the polytopes of relevance up to 1-loop scattering in detail. Our discussion will be elementary and entirely self-contained, and we will not assume prior knowledge about cluster algebras, though we will state without proof, and use, some standard cluster-theoretic results. 

\section{Causal Diamonds, the Wave Equation and Associahedra in Kinematic Space} 
To begin with, let us describe the kinematical space of relevance to tree-level scattering. We will not specify the number of dimensions of spacetime, so the kinematical space of interest is just the space of all Mandelstam invariants 
$s_{ij} = 2\,p_i \cdot p_j$. The $s_{ij}$ are of course not independent, since momentum conservation $\sum p_j^\mu = 0$ implies that $\sum_j s_{ij} = 0$. The space of independent kinematical invariants is thus $n(n-1)/2 - n = n(n-3)/2$ dimensional. A more canonical basis for all the Mandelstam invariants is given by all the propagators $X_{ij} = (p_i + \cdots p_{j-1})^2$ occurring in the planar diagrams. Note that $X_{ii+1} = 0$, so there are $n(n-1)/2 - n$ $X_{ij}$ with $i$ not adjacent to $j$. So the $X_{ij}$ are not only the propagators appearing in planar graphs, they also give a basis for all Mandelstam invariants. Note also that we have $X_{ij} = X_{ji}$. 

\begin{figure}[h!]
    \centering
    \includegraphics[scale=.75,trim={0cm .2cm 0cm 0cm}]{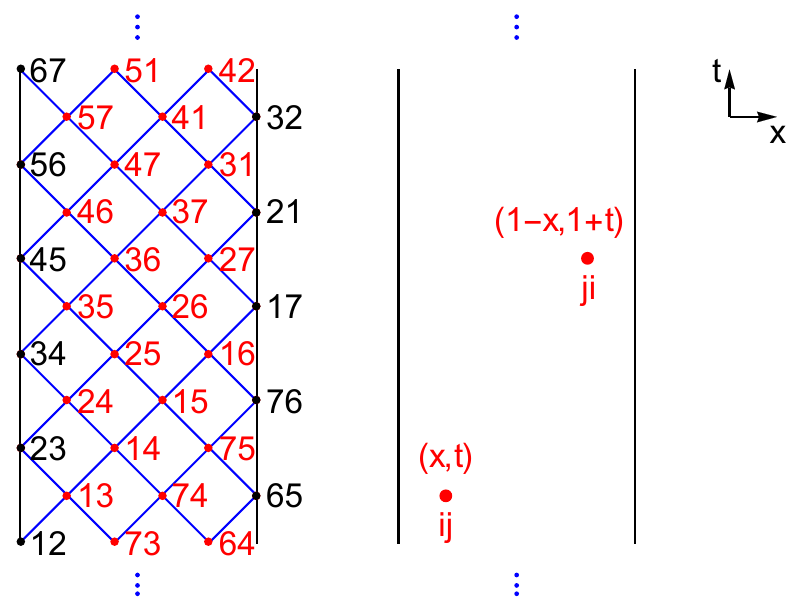}
    \caption{\label{fig:longStrip} The infinite array of planar variables.}
\end{figure}
\FloatBarrier

Our kinematic space is thus just labelled by $X_{ij}=X_{ji}$, and also enjoys a cyclic symmetry under the shift of indices $i \to i+n$.  As is familiar, we can think of the momenta $p_i$ as $i$-th edge of an $n-$gon (between vertices $i$ and $i{+}1$), then $X_{ij}$ are just the (squared) distances between the vertices $i$ and $j$ of this $n-$gon. 

It is useful to arrange all these variables $X_{ij}$, with the action of the cyclic symmetry manifest, on a two-dimensional grid in the shape of an infinite strip, as in the left part of Figure \ref{fig:longStrip}.  We will orient the grid so that lines of increasing $i,j$ run up 45 degree lines, and it will also be useful to keep the $X_{ii+1},X_{i+1 i}$ on the grid. We set $X_{ii+1}, X_{i+1 i} \to 0$, which is just the on-shell condition $p_i^2 \to 0$, and can think of these as a boundary condition on the grid. We can clearly take a ``continuum limit" which describes the kinematics of particle scattering for all possible $n$ in the same picture, simply by drawing finer grids on the same two dimensional space, { as in the right part of Figure \ref{fig:longStrip}} .Note that the infinite grid allows us to manifest the cyclic symmetry, but the $X_{ij} = X_{ji}$ requirement is a Mobius identification on the strip. 

\begin{figure}[h!]
    \centering
    \includegraphics[scale=1]{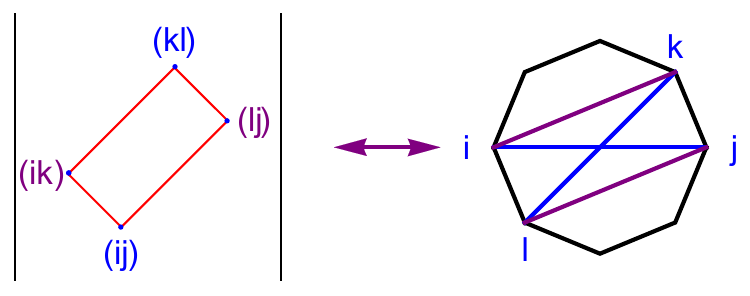}
    \caption{\label{fig:diamondNgon} Diagonals on an n-gon and the corresponding points in the $(1+1)$-dimensional space-time. }
\end{figure}
\FloatBarrier

The way we have chosen to draw the grid, as well as the labelling of the directions ``$t,x$" in the figure, suggests a causal $(1+1)$-dimensional structure in kinematic space. We will soon see the full force of this connection, but to begin with we already have a combinatorial notion of ``incompatibility'' which can be expressed equivalently in terms of Feynman diagrams, triangulations of a polygon, or causal diamonds in the spacetime: incompatible propagators are those that never appear in the same Feynman diagram, incompatible diagonals $(ij)$, $(kl)$ are diagonals which cross, and incompatible points are points in the grid which can be thought of as past and future corners of a causal diamond that fits in the spacetime, { see Figure \ref{fig:diamondNgon}}.

\begin{figure}[h!]
    \centering
    \includegraphics[scale=.5, trim={0cm 0cm 0cm 0cm}]{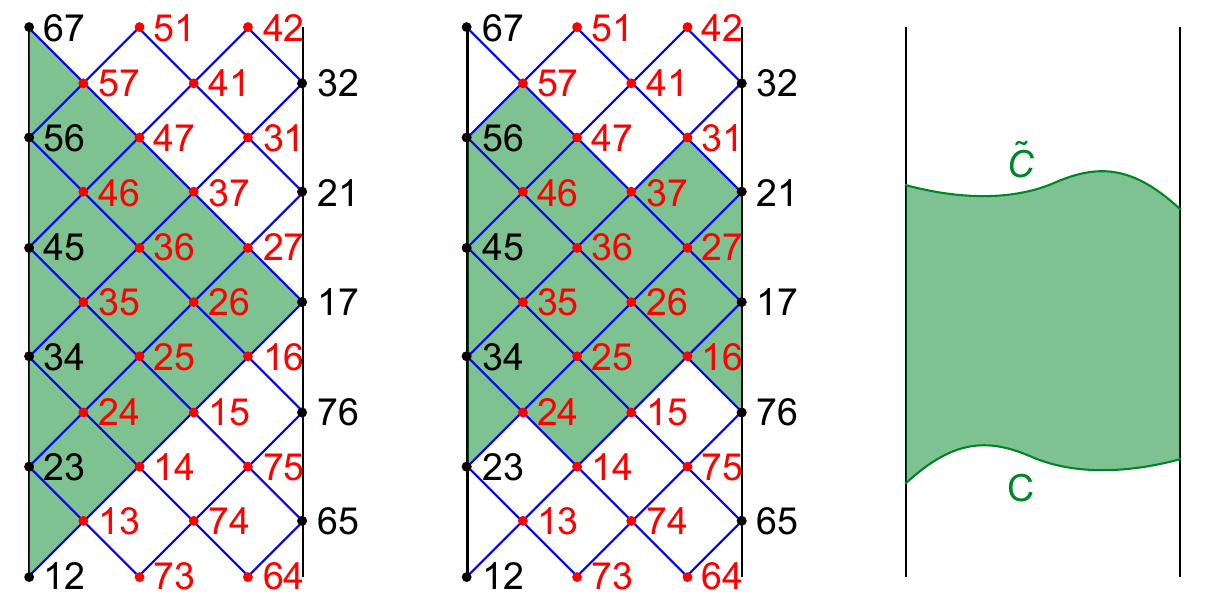}
    \caption{\label{fig:abhyregions} Examples of minimal non-redundant regions.}
\end{figure}

Note that in order to choose a non-redundant set of $X$'s for the kinematical space, {\it i.e.} in order to pick out some region that covers all $X_{ij}$'s but does not redundantly include both $X_{ij}$ and $X_{ji}$, forces us to break the cyclic symmetry in some way. We illustrate a few ways of making such a choice in Figure \ref{fig:abhyregions}. 
In general, in the continuum, we can take any region bounded by curves $C, \tilde{C}$, where $\tilde{C}$ is the image of ${\cal C}$ under the transformation $(x,t) \to (1-x, 1+t)$. 
\FloatBarrier

Having defined our kinematical space, we follow the general philosophy of \cite{Arkani-Hamed:2017vfh, Arkani-Hamed:2017mur}, and look for a natural question to be asked in this space, that will bring a positive geometry to life on (a family of) subspaces. In this view amplitudes are most fundamentally a differential form in kinematic space, fully determined by matching the canonical form of the positive geometry found on the subspace. 

\begin{figure}[h!]
    \centering
    \includegraphics[scale=.4]{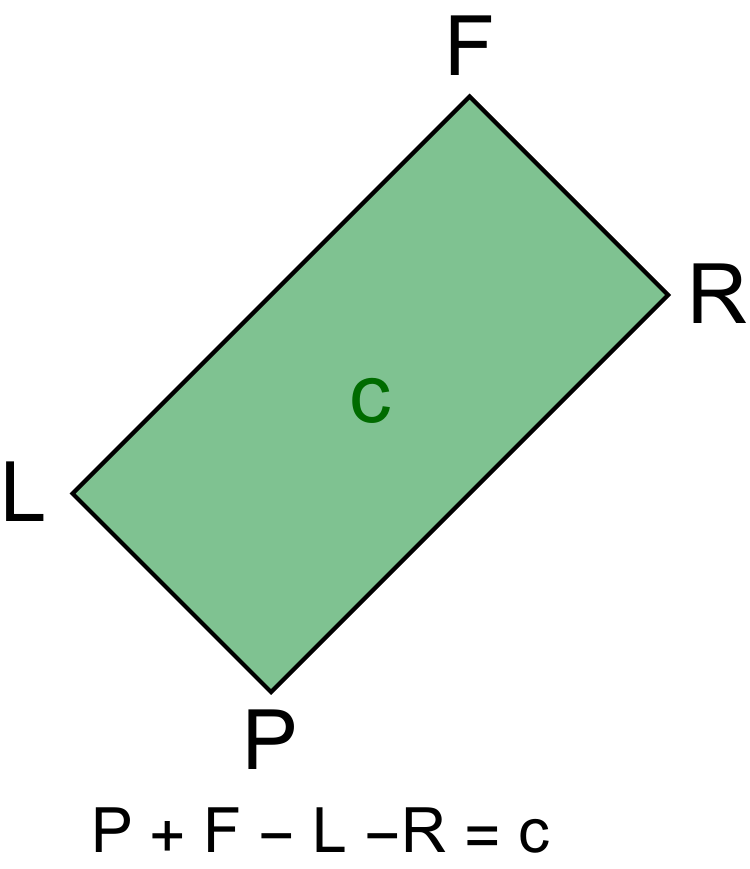}
    \caption{\label{fig:diamondRelations} A causal diamond is associated to a ``Gauss law" involving variables sitting at its corners}
\end{figure}

The question we ask is the most familiar ``dynamical" one we are used to asking, when given a causal structure in $(1+1)$ dimensions. We simply look at solutions of the wave equation, with a source, 
\begin{equation}
(\partial^2_t - \partial^2_x) X(x,t) = c (x,t) 
\end{equation}
As usual we will find it more natural to work with the lightcone co-ordinates $u =1/2 (t+x), v=1/2 (t-x)$
\begin{equation}
\partial_u \partial_v X(u,v) = c(u,v)
\end{equation}
We will make extensive use of the Gauss law for the wave equation. { For any causal diamond like the one of Figure \ref{fig:diamondRelations}}, with corners at points P(ast),L(eft),F(uture),R(ight), we have that
\begin{equation}
X_P+ X_F - X_L - X_R = C
\end{equation}
where $C$ is total ``charge" obtained by integrating the source inside the diamond. This Gauss law is the ``integral form" of the wave equation; for infinitesimally small diamonds it reduces to the wave equation. 

\FloatBarrier

\begin{figure}[h!]
    \centering
    \includegraphics[scale=0.6]{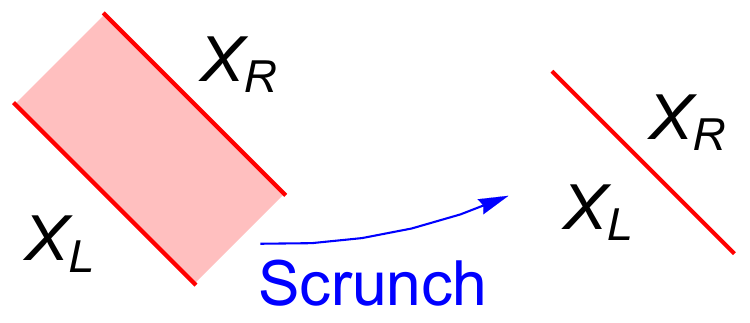}
    \caption{\label{fig:Scrunch} Scrunching a region of $(1+1)$-dimensional spacetime into a light-like line}
\end{figure}

We will use one more simple fact: given a solution of the wave equation, we can always find another solution by ``scrunching" any region bounded by parallel light rays, { as shown in Figure \ref{fig:Scrunch}}. We must merely modify the equation to account for the charge in the region to be scrunched, which must be included as a $\delta$ function source along the scrunched light-like direction. This follows trivially from the Gauss law. 
\FloatBarrier

\begin{figure}[h!]
    \centering
    \includegraphics[scale=0.4]{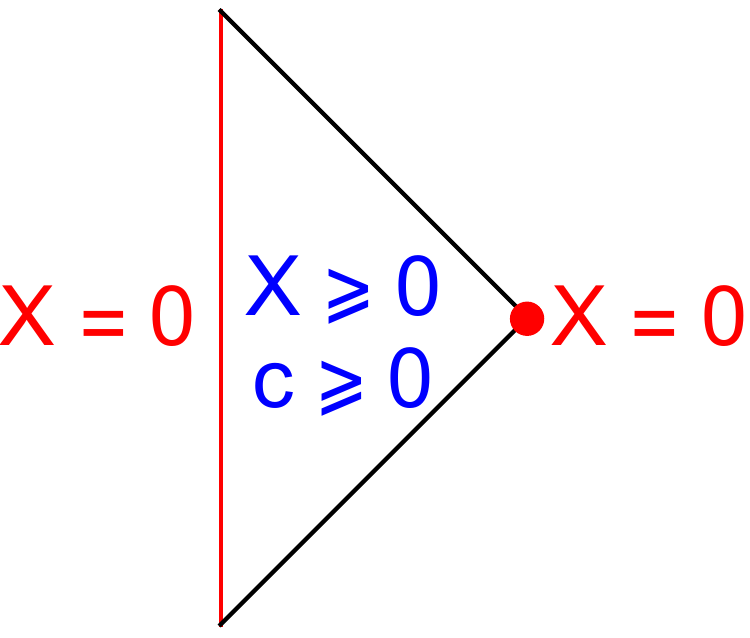}
    \caption{\label{fig:triangle} The right-triangle as a domain for the wave equation together with its boundary conditions.}
\end{figure}

Let us now look at the wave equation inside a right triangle that is half of a square-shaped causal diamond, as shown in Figure \ref{fig:triangle}. We will also inject ``positivity" into the discussion, by looking for solutions to the wave equation where $X$ is positive in interior of the triangle. To define positivity, we demand that $X$ vanishes on the tip of the triangle as well as on the space-like edge. Note that with vanishing boundary conditions, specifying $X(u) = P(u)$ on the past boundary fully specifies $X(u,v)$ in the interior of the triangle as can be seen from Gauss law. We also demand that the source $c$ is positive. 

We would thus like to find the constraints that must be imposed on the boundary function $P(u)$ in order to be compatible with positivity of $X(u,v)$ in the interior of the triangle. At first this looks to be a difficult problem, asking us to carve out some allowed region in the infinite-dimensional space of all functions of $u$. So let us simplify it, by trying to understand the constraints imposed on $P(u)$, only at a finite number of points on the boundary $P_i = P(u_i)$ on the past boundary. 

\begin{figure}[h!]
    \centering
    \includegraphics[scale=0.5]{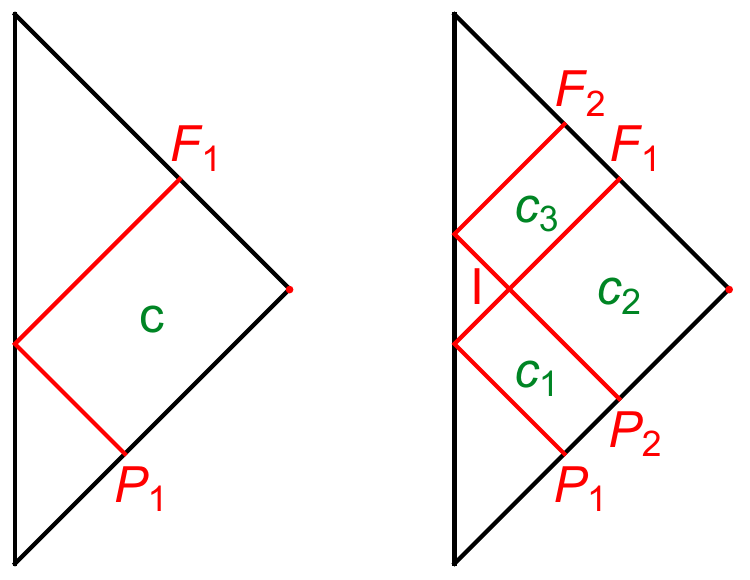}
    \caption{\label{fig:a2a3abhy} Discrete versions of the wave equation with one (resp. two) points on the past boundary shown on the left (resp. right).}
\end{figure}

Consider first the case of a single point $P_1$, as in Figure \ref{fig:a2a3abhy}. Sending out light rays associates this with the point $F_1$. Remembering that $X=0$ on the boundaries, we can use Gauss law $P_1 + F_1 = c$ to determine $F_1 = c - P_1$. Demanding that $P_1,F_1 \geq 0$ then tells us that we must have $P_1$ in the interval $0 \leq P_1 \leq c$. We next look at the case with two points $P_1,P_2$ on the past boundary. 
\FloatBarrier

\begin{figure}[h!]
    \centering
    \includegraphics[scale=0.65]{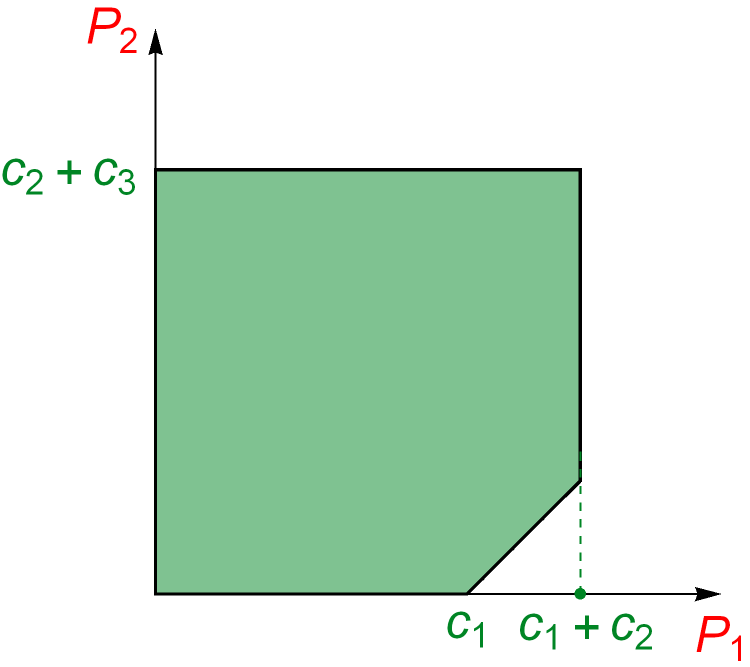}
    \caption{\label{fig:pentagona2} The positive region for the wave equation with two points on the past boundary.}
\end{figure}

Now the associated light rays intersect at a point $I$ in the interior of the spacetime. Let us now look at the Gauss laws associated with the smallest diamonds in the picture. For obvious reasons we will be referring to the Gauss laws as ``mesh relations", given in this example by
\begin{equation}
P_1 + I - P_2 = c_1, P_2 + F_1 - I = c_2, I + F_2 - F_1 = c_3\,.
\end{equation} 
These three equations can be used to solve for $I,F_1,F_2$ in terms of the initial data $P_1,P_2$, giving 
\begin{equation}
F_1 = c_1 + c_2 - P_1, F_2 = c_2 + c_3 - P_2, I = c_1 + P_2 - P_1\,.
\end{equation}
Note that quite nicely, the spacetime picture also immediately gives us these solutions; indeed these equations are just the Gauss law for the obvious diamonds that can directly used to determine $F_1,F_2,I$ from $P_1,P_2$. 
We now demand that $P_{1,2} \geq 0$, and also that $F_{1,2}, I \geq 0$. These five inequalities cuts out a region in the $P_{1,2}$ plane that is a pentagon,  as shown in Fig. \ref{fig:pentagona2}. 
\FloatBarrier

Already in this simple example we can note something slightly remarkable. The mere fact that we have $5$ inequalities does not a priori guarantee that the final shape is a pentagon. Indeed, if we replaced $c_1 + c_2, c_2+ c_3,c_1$ in the equations for $F_{1,2},I$ with general constants, we could get different shapes, since some of the inequalities may be implied by others. But the positivity of $c_{1,2,3}$, and the specific way they enter these inequalities, guarantee that we always get a pentagon for any positive $c$'s. 

Moving on this way, we discover that the region in $P_i$ space, compatible with positivity, is remarkably precisely the ABHY associahedron of \cite{Arkani-Hamed:2017mur}. Indeed, switching to the $X_{ij}$ notation for labelling points in the kinematic spacetime, we see that the past boundary of our triangle is 
covered by $X_{1,3}, \cdots, X_{1,n-1}$, and we demand the positivity of all $X_{ij} \geq 0$ on the support of the wave equation inside the triangle. This enforces for $1 \leq i  < j-1 \leq n-2$
\begin{equation}
X_{ij} + X_{i+1 j+1} - X_{i j+1} - X_{i+1 j}=c_{ij} \,.\end{equation}

Finally,  the scattering form $\Omega$ is an $(n-3)$ form on $X_{ij}$ space, completely determined by the property that, when pulled back to the subspace given by positive solutions of the wave equation, it yields the canonical form for the resulting 
associahedron. The amplitude $m_n(X)$  itself is then obtained as $\Omega = (d X_{13} \cdots dX_{1n-3}) \times  m_n(X)$. 

We have observed that positive solutions of the wave equation in kinematic space produce the ABHY associahedron, but why did this happen? We would now like to understand more deeply why this simple model gives us a polytope that factorizes on the boundary to the product of smaller polytopes of the same type. 

\begin{figure}[h!]
    \centering
    \includegraphics[scale=0.5]{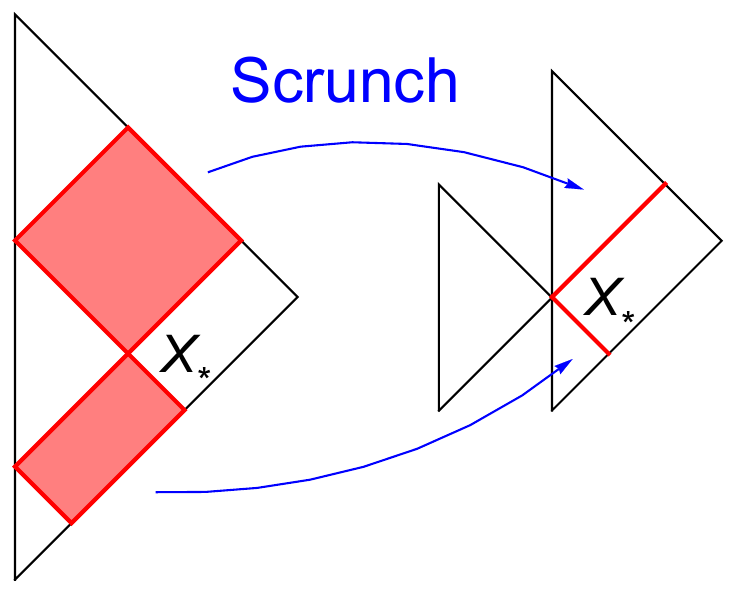}
    \caption{\label{fig:factorizingTriangles} Factorization of a right triangle domain for the wave equation into smaller domains.}
\end{figure}

It is easy to see this already in the continuum limit. Suppose we go to a boundary where we set $X_* \to 0$ at some particular point $(u_*,v_*)$. It is now natural to ask, where else can we also set $X\to 0$? This question has a beautiful answer reflecting the causal structure of our kinematic space. We can {\it not} set $X$ to zero at any pair of time-like separated points $X_P,X_F$,  which are corners of a causal diamond that fits in the spacetime. That would give a contradiction, since by using the Gauss law associated with the diamond, we would have
\begin{equation}
0<C=X_P + X_F -X_L - X_R = -X_L - X_R<0. 
\end{equation}
Therefore, having set $X_* \to 0$, there is a region bounded by the light rays emanating from $*$, where $X$ can {\it not} be set to zero, shaded red in the figure. Note we can always set a pair of space-like separated points to zero, and also time-like separated points that are far enough so there are no causal diamonds that fit in the spacetime connecting them. Note further that using the Gauss law, we can reconstruct all the $X$'s inside the shaded region, from the knowledge of $X$ on the boundaries of the unshaded region. Thus, the solution space of all $X$'s having set $X_* \to 0$, is entirely captured by ``scrunching" away the shaded regions. But after the scrunching, we see a spacetime that has factorized into the direct product of two smaller right triangles, with exactly the same boundary conditions on each smaller triangle as on the larger one, { see Figure \ref{fig:factorizingTriangles}}. This is factorization! 

\begin{figure}[h!]
    \centering
    \includegraphics[scale=0.6]{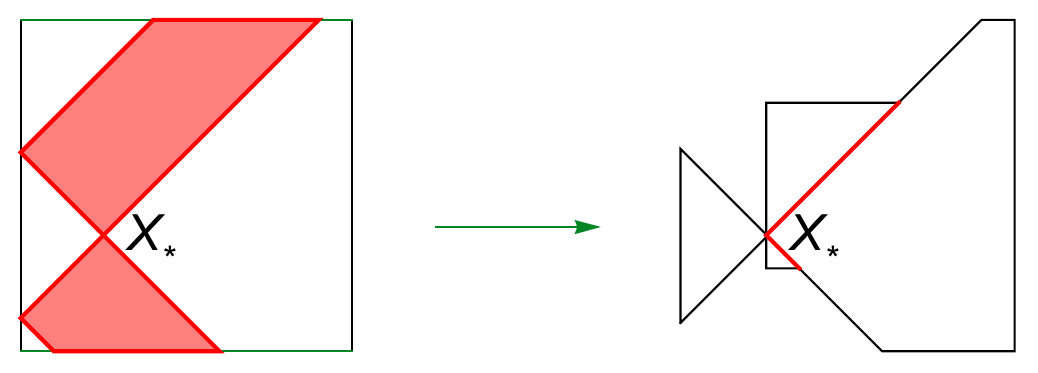}
    \caption{\label{fig:squareFactorization} Factorization of a square domain for the wave equation into smaller domains.}
\end{figure}

This argument works not just for right triangles, but for any regions of the general form we described above, bounded by curves $C, \tilde{C}$. When we set $X_* \to 0$ at any point inside the spacetime, scrunching away the regions causally connected to this point yields the direct product of two smaller spaces of the same form, bounded by new curves $C_L, \tilde{C}_L$ and $C_R, \tilde{C}_R$. We illustrate this with an example where the original spacetime is a square in Figure \ref{fig:squareFactorization}. 
The left-hand factorization is into a spacetime of the same ``right triangle'' form we were looking at previously, while the right-hand one is bounded by more interesting curves $C_R$, $\tilde C_R$, but they still define a maximal irredundant region in a smaller strip. Note again that all the $X$'s in the shaded region can be unambiguously computed using the Gauss law from $X$'s on the boundary of the region of spacetime we have kept after scrunching.

\begin{figure}[h!]
    \centering
    \includegraphics[scale=0.4]{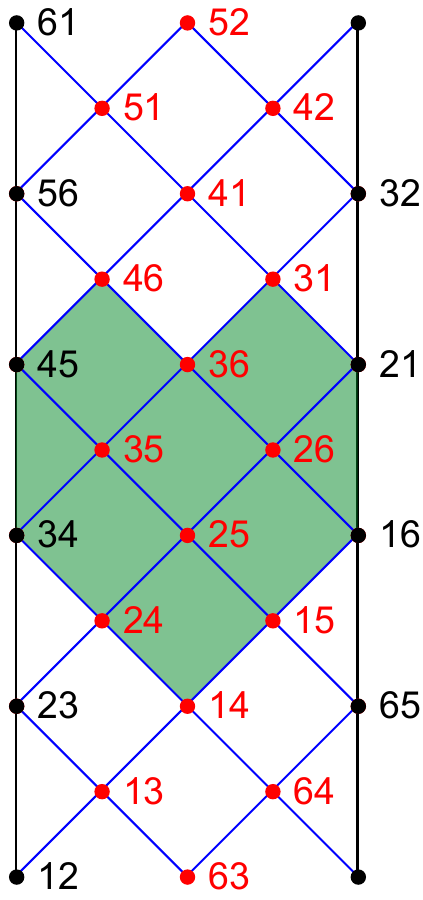}
    \caption{\label{fig:A3fig1} A domain for the wave equation yielding an $\mathcal{A}_3$ associahedron.}
\end{figure}

Thus, as shown in \cite{bazier2018abhy}, we can realize the associahedron in many different spacetimes, and not just the right triangles associated with the original ABHY construction. We simply need to take a chunk out of the discrete grid which covers all $X_{ij}$ exactly once. Another example for $n=6$ is shown in Figure \ref{fig:A3fig1}. 

\FloatBarrier
The corresponding subspace is defined by the following six mesh relations 
\begin{align}
& X_{24}+X_{35}-X_{25}=c_{24}\,, \quad \& \quad i \to i{+}1\,,\\
& X_{15}+X_{26}-X_{25}=c_{15}\,,\quad \& \quad i \to i{+}1\,,\nonumber\\
&  X_{14}+X_{25}-X_{15}-X_{24}=c_{14}\,,\quad \& \quad i \to i{+}1\,,\nonumber
\end{align}
which for all $X_{ij}>0$ (with all $C$'s positive) indeed produces a different realization of the associahedron. 

\section{Cluster Polytopes from Time Evolution as Walks on Quivers}

\begin{figure}[h!]
    \centering
    \includegraphics[scale=0.65]{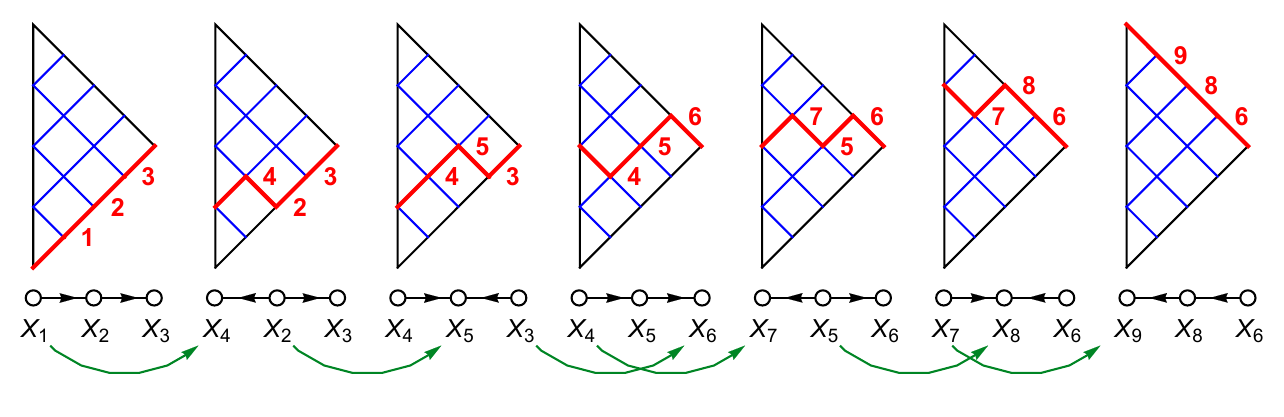}
    \caption{\label{fig:walkfig} Time evolution in $(1+1)$-dimensional spacetime and mutations on the quiver.}
\end{figure}

We will now use the ``kinematical spacetime" picture to describe the generalization of ABHY associahedra to all cluster polytopes, given in \cite{bazier2018abhy}. The motivation is a very simple and natural one. Let us return to our grid in the right triangle, and describe how we can solve the wave equation one step at a time,"via ``time evolution": beginning from the past boundary, we find some new point to the future for which $X$ can be determined by using a mesh relation-- with a single $c$ variable-- giving a new slice through the spacetime, and continuing in this way to solve for all the $X$'s in the spacetime. We illustrate this process for an example in Figure \ref{fig:walkfig}, where we begin with $3$ points $P_1,P_2,P_3$ on the past boundary. 

Now, we can capture both the shape of the slices through the spacetime, as well as the time evolution rule, in a simple and beautiful way. We draw a quiver associated with the three non-trivial $X$ variables on any slice, with and arrow from $a \to b$ if $b$ is to the future of $a$. Time evolution just corresponds to taking a node $v$ that is a source, with all arrows outgoing, and making a new quiver by ``mutation", flipping all the arrows at that node into incoming ones. We define a new variable $X_v^\prime$ via
\begin{equation}
X_v^\prime + X_v - \sum_{w \leftarrow v} X_w = c_v\,.
\end{equation}
We call this process ``walking". 

In fact we can describe everything about the time evolution in terms of walking, without having to directly reference the spacetime picture. To do so we only have to address two related questions. First, in general at any step in the process, there may be more than one node on which we can perform the mutation. Are there any restrictions on where we can mutate? For instance, in our example with $3$ points on the boundary, in the penultimate step we chose to mutate at node $7$, but if we only look at the quiver we could have also mutated at node $6$. This is clearly not sensible from spacetime evolution picture, since that would take us outside the right triangle. And second: how do we know when to stop walking? 
\begin{figure}[h!]
    \centering
    \includegraphics[scale=0.75]{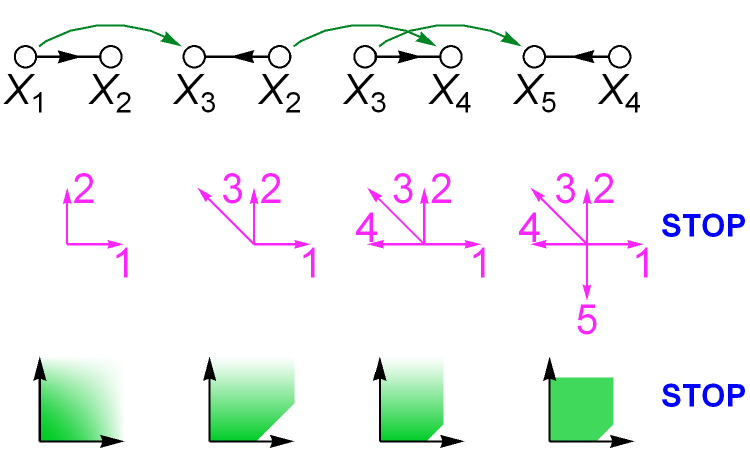}
    \caption{\label{fig:stopfig} As we walk on the quiver (first row) we collect the normal vectors (second row) to the facets of the polyhedral regions which eventually produce the associahedron (third row).}
\end{figure}
\FloatBarrier
The answer to both questions is given by thinking about how the final polytope is constructed one step at a time, as we walk. Since we are solving the $X$ variables in a sequence given by walking, we can express any variable $X_v$ we encounter in the walk as a linear expression in the variables of the initial quiver $X^{in}_i$, as $X_v = (N_v)^i X_i^{in} + C_v$ where $C$ is some linear combination of the $c$ variables seen in the walk. Since we are imposing that all the $X$'s are positive, the $(N_v)^i$, thought of as a vector in the space of the initial $X^{in}_i$, have a simple interpretation: they are the normal vectors to the facet of the polytope associated with $X_v \geq 0$. Thus as we walk, we are carving out the polytope. To begin with, we have an infinite orthant just given by  the positivity of all the $X^{in}_i$, and as we walk, we gradually chop this infinite region to smaller ones. 

The rule for what mutations are allowed, that also tells us when we have to stop walking, is then simply the following: we can mutate at any source, such that the new vector $N_v^\prime$ associated with $X_v^\prime$, lies {\it outside} the convex hull of all the $N_v$'s that have come before it. If in the process of walking we eventually reach a time where the set of all $N_v$'s cover all of space, we are forced to stop, and at this moment, the collection of $N_v$'s also give us the normals to the facets of a finite polytope. An example of this  for the case of ${\cal A}_2$ is shown in Figure \ref{fig:stopfig}.

\begin{figure}[h!]
    \centering
    \includegraphics[scale=0.75]{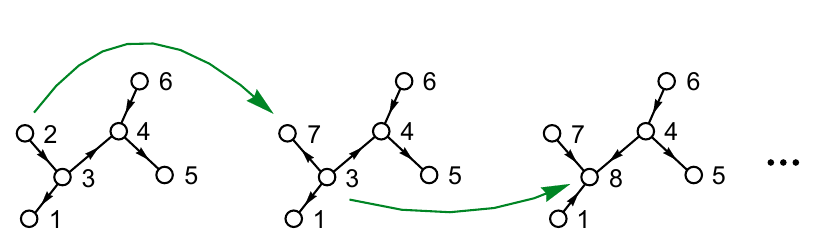}
    \caption{\label{fig:walking} Walking on a general quiver }
\end{figure}

We can now abstract away this picture of time evolution as ``walking" to any quiver. Consider any quiver, which is a tree, or more generally one with no {oriented} loops. 
We will associate variables $X_v$ with all the vertices $v$ of the quiver. There is always at least one vertex which is a ``source", with all outgoing arrows. We now define a ``walk" in exactly the same way we did above: pick a vertex that is a source, and ``mutate" to get a new quiver, by reversing the direction of all arrows from $v$, {see Figure \ref{fig:walking}}.
\FloatBarrier

We also define a new variable $X_v^\prime$ associated with the vertex $v$ in the new quiver, via
\begin{equation}
X_v^\prime + X_v - \sum_{w \leftarrow  v} n_{v,w} X_w  = c_v
\end{equation}
where we allow a slight generalization by allowing constants $n_{v,w}$, which only depend on the nodes of the quiver and not on the arrow orientations. 

We can now ask: for which quivers does the process of ``walking" stop, giving us a finite polytope associated with setting all the $X_v \geq 0$? This turns out to be extremely constraining. Experimenting with small examples easily shows that typically, the process never ends and we do not get a finite polytope. But very strikingly, finite polytopes {\it do} arise from quivers that are orientations of Dynkin diagrams! For simply laced Dynkin diagrams, the constants $n_{v,w} = 1$, while they can include integers larger than one in the non-simply laced cases. The polytopes arrived at in this way are precisely those attached to finite-type cluster algebras, described in  \cite{bazier2018abhy}. In fact, it can be shown, using results from \cite{speyer2009powers,reading2011sortable}, that for non-Dynkin quivers, the process does not end. 

It is remarkable that Dynkin diagrams arise in this setting. This connection also allows us to give a beautiful interpretation of a ``factorizing" structure of their facets: any facet of a cluster polytope, is the direct product of smaller cluster polytopes obtained by removing a node from the Dynkin diagram. The Dynkin diagram for ${\cal A}$ is a chain and so the familiar factorization into two pieces simply reflects how a chain splits into two when a single node is removed. As we will now see, the analog for type ${\cal B}/{\cal C},{\cal D}$ will allow us to connect the polytopes with scattering amplitudes through one loop in a natural way. 

\section{ABCD's of Amplitudes Through One Loop}

\begin{figure}[h!]
    \centering
    \includegraphics[scale=0.45]{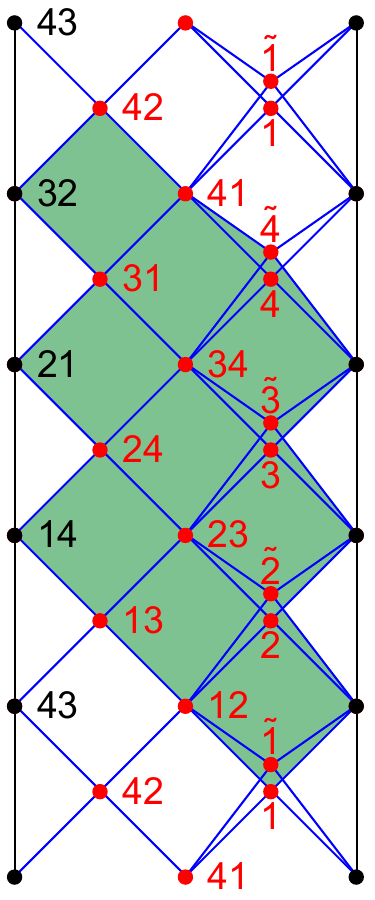}
    \caption{\label{fig:d4region} A kinematic spacetime region for $\mathcal{D}_4$.}
\end{figure}

Let us begin our discussion of the classical Dynkin diagrams with the type ${\cal D}_n$ case, which we will shortly see describes $n$-point amplitudes at 1-loop.  It turns out that for any orientation of the Dynkin quiver we have to walk $n^2 - n$ steps before being forced to stop, giving us a total of $n^2$ variables. For simplicity we will illustrate what we get from the walk, beginning with an orientation of the ${\cal D}_n$ Dynkin diagram with arrows all pointing in the same direction. This is the analog of the quiver giving us the ``right triangle" spacetime in the type ${\cal A}$. We can represent all the relations encountered on the walk by a $(1+1)$ dimensional grid  just as in type ${\cal A}$. The $n^2$ variables can be labelled on the grid as follows. We have $X_{ij}$, where we now distinguish between $X_{ij}$ and $X_{ji}$. We also have variables $X_{ii+1}$.  Finally we have {\it two} sets of variables $Y_i, \tilde{Y}_i$, associated with the antennae of the ${\cal D}_n$ Dynkin diagrams. An example of the grid for ${\cal D}_4$ (with $16$ variables and $12$ mesh relations) is shown in Figure \ref{fig:d4region}. 

\FloatBarrier

\begin{figure}[h!]
    \centering
    \includegraphics[scale=0.75]{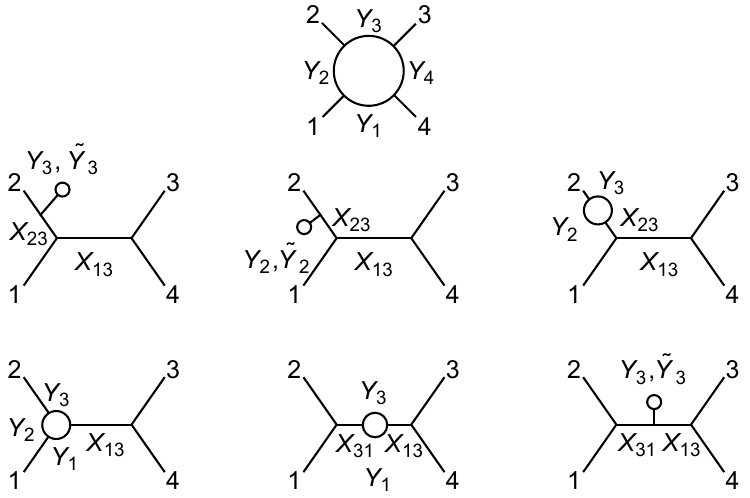}
    \caption{\label{fig:graphsTogether} Labelling of cluster variables by Feynman propagators.}
\end{figure}

These variables are naturally associated with propagators of 1-loop graphs. We distinguish $X_{ij}$ from $X_{ji}$ since the internal propagator outside the loop, can be thought of going around the loop in one orientation or the other. The variables $X_{ii+1}$ are thought as attached to the ``internal" propagator associated with external bubbles. Finally, $Y_i$ is associated with the loop itself. The only peculiarity is the presence of both $Y_i$ and $\tilde{Y}_i$. This has to do with tadpole diagrams: each tadpole diagram is associated both with a loop variables and the tadpole propagator, and the ${\cal D}_n$ polytope treats this symmetrically, so such a diagram has both propagators $Y_i, \tilde{Y}_i$. We illustrate how these variables are attached to the 1-loop graphs in the case of ${\cal D}_4$ in Figure  \ref{fig:graphsTogether}, note that everywhere we have $Y_i$ there is also a graph where all the $Y_i$ are replaced by $\tilde{Y}_i$, but the only graphs where they occur together are the tadpoles. We will soon describe a smaller polytope that avoids this doubling. 

From the walk for the grid for ${\cal D}_n$, we read off the mesh relations. We have for $1\leq i\leq n{-}1$
\begin{align}
&Y_i + Y_{i+1} - X_{ii+1} = c_{i}, \quad \tilde{Y}_i + \tilde{Y}_{i+1} - X_{ii+1} = \tilde{c}_{i},
\nonumber \\
&X_{ii+1} + X_{i+1 i+2} -X_{i i+2} - Y_{i+1} - \tilde{Y}_{i+1} = c_{ii+1}, 
\end{align}
and for non-adjacent $i,j \neq n$, and $1<i<n{-}1$, $j=n$ 
\begin{equation}
X_{ij} + X_{i+1 j+1} - X_{i+1 j} - X_{i j+1} = c_{ij}\,.
\end{equation}
Requiring positivity for all the variables, these equations cut out the $\mathcal{D}_n$ polytope. Note that, especially clearly at large $n$, these equations are just the wave equation with some extra degrees of freedom (doubling of $Y,\tilde{Y}$) on one boundary of the spacetime. 

Starting with these equations, it is easy to determine the boundary structure of the ${\cal D}_n$ polytope. For $X_{ij}=0$ with non-adjacent $i,j$, the facet factorizes into ${\cal A} \times {\cal D}$. To be precise, we have two cases here:
\be
{\cal D}_n \xrightarrow[]{\partial_{X_{ij}}} \begin{cases}
&{\cal A}_{i{-}j{-}2} \times {\cal D}_{n{+}j{-}i{+}1}\,, \quad {\rm for}~j<i{-}1\,,\\
&{\cal A}_{n{+}i{-}j{-}2} \times {\cal D}_{j{-}i{+}1}\,,\quad {\rm for}~i<j{-}1\,.
\end{cases}
\ee
This has a clear interpretation of a 1-loop integrand factorizing into a lower point 1-loop integrand times a tree-level amplitude. With $j=i{-}2$ the tree-level factor corresponds to a $3$-point amplitude, which is just a constant, and accordingly we have ${\cal D}_{n{-}1} \times {\cal A}_0\simeq {\cal D}_{n{-}1}$. For $i=j{-}2$ the 1-loop factor corresponds to a $3$-point integrand, which is given by the non-trivial ${\cal D}_3$, thus we have a boundary of the form ${\cal A}_{n{-}4} \times {\cal D}_3$. 
 
For $X_{i, i{+}1}=0$, the facet factorizes into a $n$-point tree, ${\cal A}_{n{-}3}$, times an external bubble, or ${\cal A}_1 \times {\cal A}_1$ (which can be though of as the degenerate case ${\cal D}_2$), thus we have
\be
{\cal D}_n \xrightarrow[]{\partial_{X_{i,i+1}}} {\cal A}_{n{-}3} \times {\cal A}_1 \times {\cal A}_1.
\ee
This is familiar from ${\cal D}_n$ Dynkin diagram, which factorizes into three terms at this special node. 
Finally, for $Y_i=0$ (or $\tilde{Y}_i=0$), as is also familiar from Dynkin diagram, the result is simply the $(n{+}2)$-point tree ${\cal A}_{n{-}1}$:
\be
{\cal D}_n \xrightarrow[]{\partial_{Y_i {\rm or} \tilde{Y}_i}} {\cal A}_{n{-}1}(i,i{+}1,\cdots, i{-}1, \pm, \mp)\,, 
\ee
We refer to this as the ``cut" facet since it can be interpreted as cutting the loop open to get the ``forward limit" of $(n{+}2)$-point tree with the additional legs labelled as $+$ and $-$ 
. For example, ${\cal D}_3$ has 3 quadrilaterals ${\cal A}_1 \times {\cal A}_1$ from the second type and 6 pentagons ${\cal A}_2$ from the third type; for ${\cal D}_4$, there are { 12 copies of ${\cal A}_3$ and 4 copies of ${\cal A}_1 \times {\cal A}_1\times {\cal A}_1$}. All these facts can be derived by studying the ABHY construction from the mesh diagram above, and it is crucial that we have various realizations of type ${\cal A}$ associahedra, since they are needed in such factorizations. 

Finally, just as for tree amplitudes, the scattering form that yields the integrand for 1-loop amplitudes, is just the unique form on the kinematic space that pulls back to the canonical form of the ${\cal D}_n$ polytope on the subspace defined by the walk. 

\begin{figure}[h!]
\minipage{0.5\textwidth}
\centering
\includegraphics[scale=0.45, trim={.5cm 2.75cm .35cm -1.5cm}]{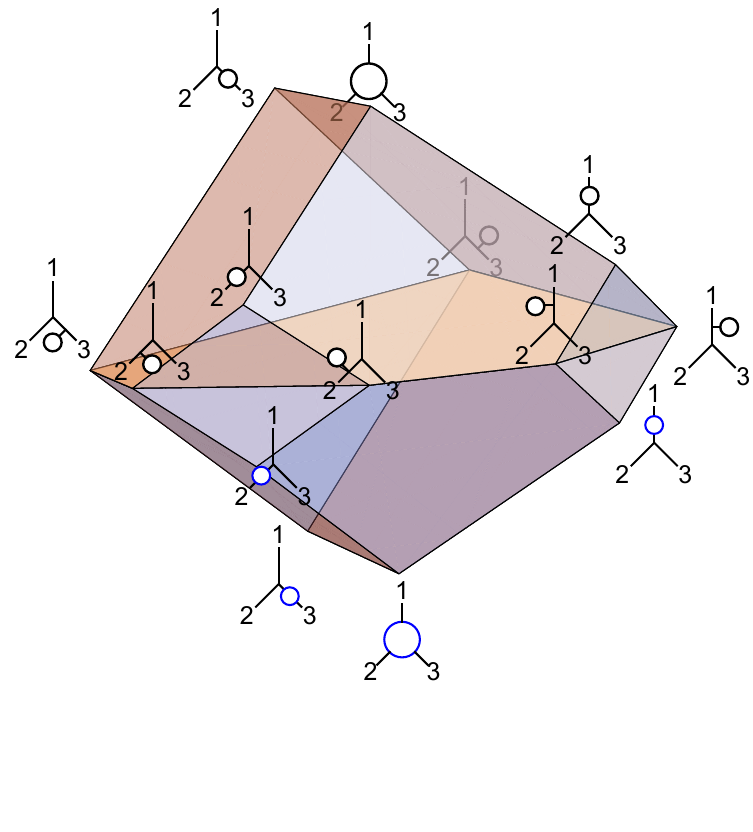}
\endminipage\hfill
\minipage{0.5\textwidth}
\centering
\hspace{-.2cm}
\includegraphics[scale=0.45, trim={.8cm 2.75cm .8cm 1.5cm}]{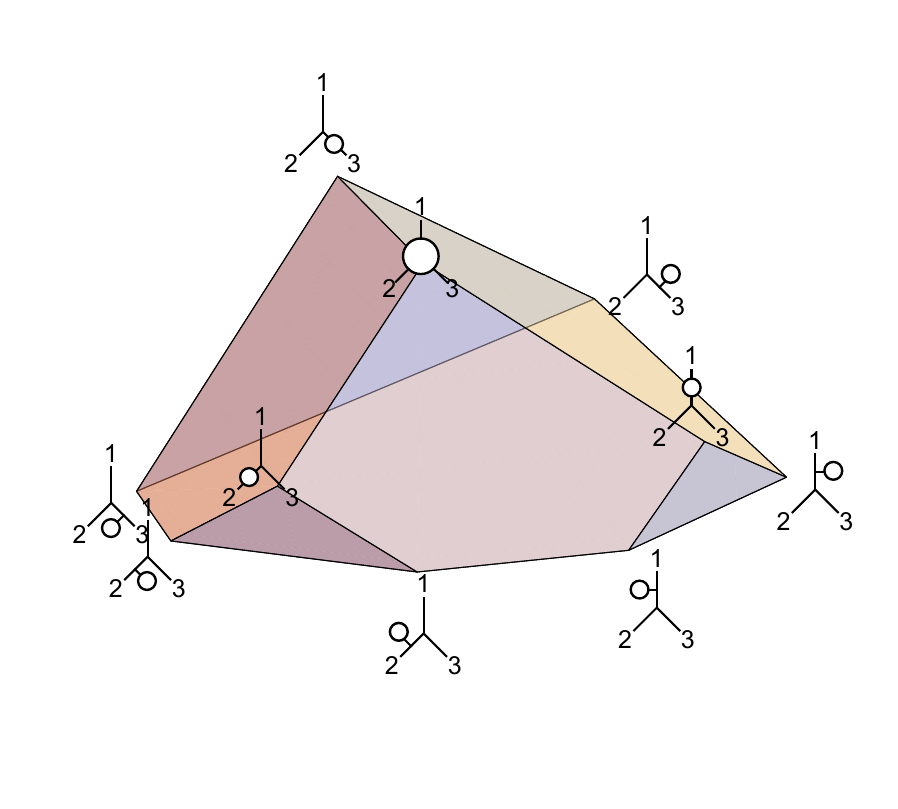}
\endminipage\hfill
\caption{\label{fig:d3d3bar} The cluster polytope $\mathcal{D}_3$ (left) and the new polytope $\overline{\mathcal{D}}_3$ (right).}
\end{figure}
\FloatBarrier 

\begin{figure}[h!]
    \centering
    \includegraphics[scale=0.75]{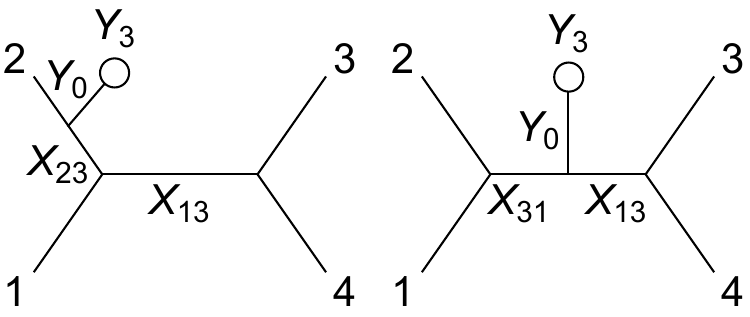}
    \caption{\label{fig:graphsYO} Tadpole diagrams and the associated variables}
\end{figure}

The ${\cal D}_n$ polytope correctly captures the combinatorics of the 1-loop factorization. But it does so in a somewhat redundant way, by doubling the number of loop variables, using both $Y_i$ and $\tilde{Y}_i$. There is a natural way to remove this un-necessary doubling, and land on a polytope that more efficiently captures the combinatorics of 1-loop diagrams. Let us begin by imposing a symmetry on the mesh constants so that $c_i = \tilde{c}_i$. It is then easy to verify that with this choice, on the solution of the equations, the differences $Y_i - \tilde{Y}_i$ are actually independent of $i$; let us set $Y_ 0 = Y_i - \tilde{Y}_i$. Now it is easy to see that all vertices of ${\cal D}_n$ which correspond to the tadpole one-loop diagrams lie on $Y_0 = 0$. The reason is simple: in any tadpole diagram, we are setting both $Y_i, \tilde{Y}_i \to 0$, and thus both terms in $Y_0 \to 0$. As we have remarked, the tadpole diagrams are the only ones in which both $Y_i$ and $\tilde{Y}_i$ appear, and so the other vertices of ${\cal D}_n$ involving just $Y_i$ have $Y_0 >0$ while the ones involving $\tilde{Y}_i$ have $Y_0<0$. Thus if we simply impose $Y_0 \geq 0$, we cut the ${\cal D}_n$ polytope in half, getting a new polytope we call $\bar{{\cal D}}_n$, { see Figure \ref{fig:d3d3bar}}. This polytope has a single new facet $Y_0$, on which all the tadpole diagrams live, so in diagrams, we can universally associate $Y_0$ with all the tadpole propagators, {see Figure \ref{fig:graphsYO}}. Thus, $\bar{{\cal D}}_n$ beautifully captures all the 1-loop planar diagrams of the bi-adjoint theory, with no ``doubling", and a natural new variable associated with tadpole diagrams. 

As we have just seen, the $Y_0=0$ facet of $\bar{\cal D}_n$ is an $(n-1)$-dimensional polytope. But setting $Y_0 \to 0 \rightarrow Y_i = \tilde{Y}_i$ is just ``folding" the antennae of the Dynkin diagram, producing the Dynkin diagram of ${\cal B}_{n-1}$! Thus we arrive at cluster polytope for ${\cal B}_{n-1}$,  also known as ``cyclohedron", which has $n^2 -n$ facets. Its ABHY realization is given just by that of ${\cal D}_n$, setting $\tilde{c}_i=c_i$ and also $Y_i = \tilde{Y}_i$. This polytope is associated with all the tadpole diagrams at 1-loop. 
\begin{figure}[h!]
    \centering
    \includegraphics[scale=0.5]{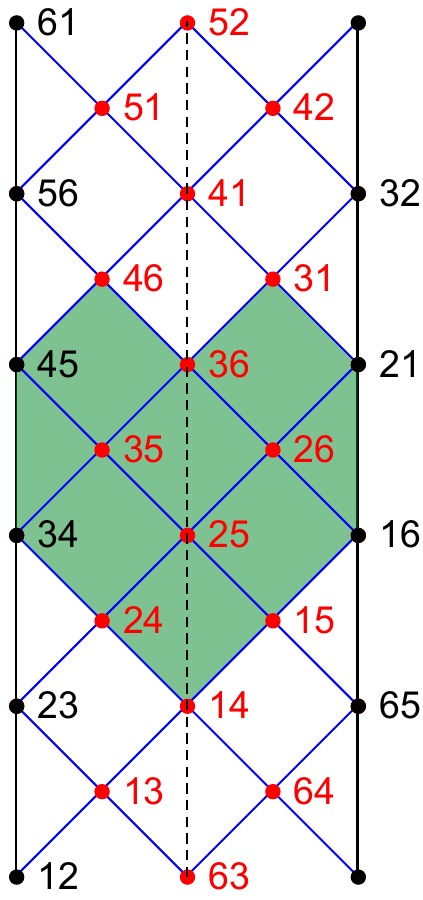}
    \caption{\label{fig:a3fig2} A region for ${\cal A}_{2n-3}$ compatible with central symmetry of the $2n$-gon.}
\end{figure}
We can similarly identify the ${\cal C}_{n-1}$ Dynkin diagram by ``folding" the Dynkin diagram of ${\cal A}_{2n -3}$ around the middle node of it's quiver. In our language, the variables of ${\cal A}_{2n-3}$ are associated with the chords $X_{ij}$ of a $2n$-gon. 
The identification leading to ${\cal C}_{n-1}$ is to restrict to ``centrally symmetric" triangulation, imposing the condition that $X_{i,j} = X_{j+n,i+n}$. We can realize this with ABHY by using a mesh for ${\cal A}_{2n-3}$ preserving this symmetry, { as in Figure \ref{fig:a3fig2}} (see also~\cite{Li:2018mnq}). 

It is very easy to see that the shape of the polytopes for ${\cal B}_{n-1}$ and ${\cal C}_{n-1}$ are exactly the same. Indeed, if we rescale $X_{ij} \to 2 X_{ij}$ and $c_{ij} \to 2 c_{ij}$ in the equations for ${\cal B}_{n-1}$, we get the equations for ${\cal C}_{n-1}$ up to a simple relabelling of the variables. 

\section{Outlook}
We have seen a remarkable connection between causal structure and wave-equation dynamics in kinematic space, and scattering amplitudes, with amplitudes interpreted as the answer to natural combinatorial and geometric questions directly in kinematic space, while the rules of locality and unitarity -- spacetime and quantum mechanics -- arise as derivative notions. In this letter we have focused on particle scattering but one can study natural ``stringy" generalizations by considering ``stringy canonical forms"~\cite{Arkani-Hamed:2019mrd}, which provide a natural definition and extension of canonical forms for polytopes, deformed by a parameter $\alpha'$. It is fascinating that when applied to ABHY associahedra, the most natural integrals reproduce the usual Koba-Nielsen string integrals, while for other finite type we have ``cluster string integrals" generalizing string amplitudes, which have factorization associated with Dynkin diagram even at finite $\alpha'$~\cite{20193}! What underpins these integrals is a completely rigid geometric realization of the combinatorics of generalized associahedra -- ``binary (positive and complex) geometries", which naturally generalize the moduli space of (open- and closed-) string worldsheets~\cite{20193}. 

We will leave a more extensive exposition and exploration of the ideas presented in this letter to future work~\cite{future}, where amongst other things we will show how the geometry of the cluster polytopes yields new expression for the amplitudes, which are strikingly more efficient than Feynman diagrams. Already in~\cite{Arkani-Hamed:2017mur}, we have seen new formulas for bi-adjoint $\phi^3$ tree amplitudes from triangulations of the ${\cal A}_{n{-}3}$ associahedron (see also \cite{He:2018svj}); similarly any triangulations of the cyclohedron or $\overline{\cal D}_n$ polytope directly lead to new one-loop formulas. But our $(1+1)$ spacetime picture of the ABHY associahedron manifests a beautiful ``self-projection" property that leads to especially powerful recursion relations. To illustrate the geometric fact, let us return to the example of right triangle kinematic spacetime. Let us consider the polytope in the space with $n$ points on the past boundary $P_1, \cdots, P_n$,  but suppose we do not care about the constraints placed on one of the variables, say $P_j$. Geometrically, we just project the polytope through the $P_j$ direction. But the spacetime picture makes it clear that the projected polytope is nothing but the $(n{-}1)$ dimensional associahedron obtained by simply removing $P_j$ to begin with (together with the obvious redefinition of the mesh constants for the Gauss law)! 

This self-projecting property allows us to give a novel sort of ``triangulation" of the associahedron, based on projecting the polytope to one of the facets, giving new, efficient recursion relations for tree and one-loop amplitudes. For example, for ${\cal A}_{n{-}3}$ the projection that corresponds to ``soft-limit" gives a recursion with $n-3$ terms only, and for ${\overline{\cal D}_n}$ the projection to the tadpole facet gives a ``forward-limit" formula for one-loop integrand (such recursions can also be derived from field-theoretical considerations~\cite{Yang:2019esm} or general properties of canonical forms of simple polytopes \cite{Salvatori:2019phs}). We content ourselves to present these formulae here, their derivation from the aforementioned projection property of cluster polytopes will be presented in~\cite{future}. For tree amplitudes we find
\begin{align}
m_n = \sum_{i=4}^{n} \left(\frac{1}{X_{1,3}} + \frac{1}{X_{2,i}}\right) \hat{m}_{n_L} \times \hat{m}_{n_R},
\label{eq:recursiontree}
\end{align}
where $m_{n_L}$ and $m_{n_R}$ are the two lower point amplitudes into which $m_n$ factorizes on the channel $X_{2,i}=0$. The hats denote a deformation of these amplitude defined by $X_{2,j} \to X_{2,j} - X_{2,i}$ for all $i \ne j$. At $4$- point \eqref{eq:recursiontree} gives $m_4 = \frac{1}{X_{1,3}} + \frac{1}{X_{2,4}},$ while at 5-point we obtain
\begin{align}
\label{eq:5ptsoft}
\hspace{-.5cm}
m_5 &=\left(\frac{1}{X_{2,4}}+\frac{1}{X_{1,3}}\right) \left(\frac{1}{X_{2,5}-X_{2,4}}+\frac{1}{X_{1,4}}\right)\nonumber\\
&+\left(\frac{1}{X_{2,5}}+\frac{1}{X_{1,3}}\right)\left(\frac{1}{X_{3,5}}+\frac{1}{X_{2,4}-X_{2,5}}\right)\,.
\end{align}
The same projection property underlying \eqref{eq:recursiontree} is generalized to all cluster polytopes, and gives a similar recursive formula for the one-loop integrand of bi-adjoint theory,
\begin{align}
m_{n}^{\mathrm{1-loop}} =&\sum_{i=2}^{n-1} \left(\frac{1}{X_{1,n-1}} + \frac{1}{X_{i,n}}\right) \left(\hat{m}_{n_L}\times \hat{m}^{\mathrm{1-loop}}_{n_R}\right) \nonumber\\
+&\sum_{i=2}^{n-1} \left(\frac{1}{X_{1,n-1}} + \frac{1}{X_{n,i}}\right) \hat{m}^{\mathrm{1-loop}}_{n_L}\times \hat{m}_{n_R} \nonumber\\
+& \left(\frac{1}{X_{1,n-1}} + \frac{1}{Y_{n}}\right)\hat{m}_{n+2},
\end{align}
where again $n_L$ and $n_R$ denote the two sets into which the $n$ particles factorize on the pole $X_{i,n}=0$ of $m_{n}$. The hats denote deformation of the corresponding tree level amplitudes and 1-loop integrands, the deformations are defined by $X_{j,n} \to X_{j,n} - X_{i,n-1}$ and $Y_{n} \to Y_{n} - X_{i,n}$.
Note that we also have a contribution from an $n+2$ tree level amplitude, which correspond to the facet $Y_{n}$ of $\overline{\mathcal{D}}_n$, its deformation is defined by $X_{j,n} \to X_{j,n} - Y_{n}$.

Furthermore, the way $\overline{\mathcal{D}}_{n}$ is constructed from $\mathcal{D}_{n}$ by slicing along the tadpole plane suggests another projection, which in turns implies the following ``forward-limit'' formula for bi-adjoint theory:
\begin{align*}
m_n^{\rm{1-loop}} = \sum_{i = 1}^n \frac{1}{Y_i} \hat{m}_{n+2}(1,\dots,i,-,+,i+1,\dots,n),
\end{align*}
where the tree level amplitudes are deformed to the forward-limit configuration $\hat{m}_{n+2}$ defined by $Y_j \to Y_j - Y_i$, up to a shift in the loop momentum this formula agrees with the forward-limit formula of~\cite{He:2015yua}.

Let us close by mentioning a few avenues of exploration for future work. One natural question is to ask whether it is possible to meaningfully talk about an ``infinite" associahedron directly in the continuum picture of the $(1+1)$ kinematical spacetime.  The ``vertices" of this infinite associahedron should correspond to various sorts of spatial slices through the spacetime, but should also have a natural fractal structure associated with infinitely many factorizations, and it would be interesting to find a naturally continuous description of this geometry. 

Another has to do with the possible physical interpretation of exceptional cluster polytopes. Obviously the cluster polytopes of classical Dynkin type have a parameter ``$n$" that allows them to be interpreted as amplituhedra for the bi-adjoint $\phi^3$ theory through one-loop order for any number of particles. The exceptional types do not have such an $n$, but perhaps they have other physical interpretations nonetheless. 

Finally, the most pressing question is the extension of these ideas to describing amplitudes at all loop orders. As we stressed in our introductory remarks, we aimed to present the results of this letter in the most elementary and self-contained way possible, without any explicit reference to cluster algebras. But the cluster-theoretic viewpoint is powerful, and instantly suggests the association between the ${\cal A}_n$/${\cal D}_n$ cluster polytopes and tree/one-loop amplitudes we have described in this letter. There are cluster algebras associated with triangulations of surfaces with any number of boundary points and internal punctures. Feynman diagrams are simply the duals of these triangulations, so if there is to be any obvious connection between amplitudes and cluster algebras, it should be in the setting of these cluster algebras associated with surfaces. Indeed the algebras associated to a disk with zero or one punctures, are just the finite-type type ${\cal A}$ and ${\cal D}$ cluster algebras~\footnote{It is interesting to note that one of the few topological types of surfaces which are not associated to a cluster algebra is an annulus with marked points on one boundary. The moduli space of this surface is associated with a polytope, called Halohedron, which has a combinatorial structure very similar to $\overline{\mathcal{D}}_n$ and for this reason was considered in \cite{Salvatori:2018aha} in connection with one-loop integrands for the bi-adjoint theory as well.}. Beyond one loop, it is natural  to hope that the amplituhedron for bi-adjoint $\phi^3$ theory at $L$-loops orders is associated with the cluster algebra of surfaces with $L$ internal pictures. The challenge is that the cluster algebras in this case have infinitely many variables not obviously seen in the amplitudes. 
We hope to see significant progress on this question in the future. 

\section*{Acknowledgments}
We thank Hadleigh Frost,  Thomas Lam, Zhenjie Li, Alexander Postnikov, Lauren Williams, Qinglin Yang and Chi Zhang for stimulating discussions. S.H., G.S. and H.T. would like to thank the hospitality of the Institute of Advanced Study, Princeton, and Center for Mathematical Sciences and Applications, Harvard, where parts of the work were done. SH’s research is supported in part by the Key Research Program of the Chinese Academy of Sciences, Grant NO. XDPB15, and by National Natural Science Foundation of China under Grant No. 11935013,11947301, 12047502,12047503. G.S. is supported by the Simons Investigator Award\#376208 of A. Volovich, and also thanks Institute of Theoretical Physics, Beijing, for hospitality, and Istituto Nazionale di Fisica Nucleare for generously providing travel funding. H.T. was supported by an NSERC Discover Grant and the Canada Research Chairs program.

\end{document}